%% file: cosmoHammer.tex
\documentclass[authoryear,12pt,5p,times]{elsarticle}

\usepackage{amssymb}
\usepackage{amsmath}
\usepackage{graphicx}
\usepackage{xspace}
\usepackage{color}
\usepackage{hyperref}
\usepackage{listings}
\usepackage{microtype}
\usepackage{array}

\journal{Astronomy and Computing}

\newcommand{\var}{\text{Var}}
\newcommand{\ch}{\texttt{CosmoHammer}\xspace}
\newcommand{\emcee}{\texttt{emcee}\xspace}
\newcommand{\cosmomc}{\texttt{CosmoMC}\xspace}
\newcommand{\camb}{\texttt{CAMB}\xspace}
\begin{document}

\begin{frontmatter}



\title{CosmoHammer:\\ Cosmological parameter estimation with the MCMC Hammer}


\author[fhnw]{Jo\"el Akeret\corref{cor1}}
\ead{jakeret@phys.ethz.ch}

\author[eth]{Sebastian Seehars}

\author[eth]{Adam Amara}

\author[eth]{Alexandre Refregier}

\author[fhnw]{Andr\'e Csillaghy}

\cortext[cor1]{Corresponding author}

\address[fhnw]{University of Applied Sciences Northwestern Switzerland, Institute of 4D Technologies, Steinackerstrasse 5, 5210 Windisch, Switzerland}
\address[eth]{ETH Zurich, Department of Physics, Wolfgang-Pauli-Strasse 27, 8093 Zurich, Switzerland}

\begin{abstract}

We study the benefits and limits of parallelised Markov chain Monte Carlo (MCMC) sampling in cosmology. MCMC methods are widely used for the estimation of cosmological parameters from a given set of observations and are typically based on the Metropolis-Hastings algorithm. Some of the required calculations can however be computationally intensive, meaning that a single long chain can take several hours or days to calculate. In practice, this can be limiting, since the MCMC process needs to be performed many times to test the impact of possible systematics and to understand the robustness of the measurements being made. To achieve greater speed through parallelisation, MCMC algorithms need to have short auto-correlation times and minimal overheads caused by tuning and burn-in. The resulting scalability is hence influenced by two factors, the MCMC overheads and the parallelisation costs. In order to efficiently distribute the MCMC sampling over thousands of cores on modern cloud computing infrastructure, we developed a Python framework called \ch which embeds \emcee, an implementation by \citet{Foreman-Mackey2012} of the affine invariant ensemble sampler by \citet{Goodman2010a}. We test the performance of \ch for cosmological parameter estimation from cosmic microwave background data. While Metropolis-Hastings is dominated by overheads, \ch is able to accelerate the sampling process from a wall time of 30 hours on a dual core notebook to 16 minutes by scaling out to 2048 cores. Such short wall times for complex data sets opens possibilities for extensive model testing and control of systematics. 

\end{abstract}

\begin{keyword}
Markov chain Monte Carlo methods \sep Cloud computing \sep Cosmological parameter estimation


\end{keyword}

\end{frontmatter}


\input{introduction}
\input{problem}
\input{solution}
\input{components}
\input{testsandresults}
\input{benchmarks}
\input{conclusion}

\appendix

\input{appendixA}
\input{appendixD}
\input{appendixB}

\section*{Acknowledgements}

We want to thank Joel Berg\'e, Lukas Gamper and Joe Zuntz for helpful discussions, as well as Laurenz Gamper for his help with licensing and homepage.

\bibliographystyle{elsarticle-harv}
\bibliography{cosmoHammer}







\end{document}

%% file: introduction.tex
\section{Introduction}

Bayesian inference is a standard procedure in cosmology when measurement results are compared to predictions of a parameter-dependent model. The likelihood function $\mathcal L(\theta)$ is defined as the conditional probability of a measurement outcome $x$ given the model parameters are fixed to $\theta$: $\mathcal L(\theta) := p(x|\theta)$. Bayesian inference tells us how to update our knowledge from a prior distribution $q(\theta)$ to a new posterior distribution $p^{new}(\theta)$ which accounts for the recent measurement:
\begin{equation*}
	p^{new}(\theta) \propto q(\theta)p(x|\theta).
\end{equation*}

For exploring the likelihood function $\mathcal L$ or the posterior distribution $p^{new}$ of the parameters, Markov chain Monte Carlo (MCMC) algorithms are today's method of choice when no functional expressions are available. Starting with the analysis of cosmic microwave background (CMB) data in 2001 \citep{Christensen2001, Knox2001}, MCMC methods turned into a vital tool in the analysis of astronomical data from all sorts of cosmological probes.

Most of the analyses in the literature rely on a particular instance of an MCMC method, the so-called Metropolis-Hastings algorithm from 1970 \citep{Metropolis1953, Hastings1970}. It is based on a random walk in the parameter space of the likelihood, serially proposing new positions that are accepted or rejected according to its likelihood weights. The Fortran program \cosmomc by \citet{Lewis2002}, for example, is a widely used and very successful tool for parameter estimation in the cosmology community, based on the Metropolis-Hastings algorithm and the Boltzmann integrator \camb (Code for Anisotropies in the Microwave Background by \citet{Lewis2000}). A large number of scientific projects have used \cosmomc, among them prominent observations as the Wilkinson Microwave Anisotropy Probe (WMAP) \citep{Dunkley2009} or the Sloan Digital Sky Survey \citep{Tegmark2004}.

Recently, \citet{Foreman-Mackey2012} presented \emcee, a Python implementation of a novel MCMC algorithm by \citet{Goodman2010a} which has several potential advantages over Metropolis-Hastings: the sampling depends on less tuning-parameters and is independent of linear transformations of the parameters. Furthermore, \emcee is not based on a single iterative random walk but uses an ensemble of walkers which can be moved in parallel.

Creating samples for the estimation of cosmological parameters from CMB measurements, for example, typically takes a few hours or even days on a desktop computer when using the Metropolis-Hastings algorithm. Whenever the sampling process has to be repeated a number of times in order to study the fit of various distinct models to the data or to find out about the systematics of a measurement, such a long run time gets increasingly problematic. We  therefore focus on the parallelisability of MCMC methods in order to minimise the wall time of the calculation.

One of the main questions we try to answer is how fast one can, in principle, generate a useful sample of a given distribution when using \emcee on an extended computing environment such as a cloud service or grid computer. For this purpose, we combined the Boltzmann integrator \camb and the WMAP likelihood code and data \citep{Larson2011, Komatsu2011} (both written in Fortran90) with the \emcee sampler by \citet{Foreman-Mackey2012} in a Python framework---called \ch in the following---allowing us to study the example of parameter inference from CMB data in detail. As parameter estimation with CMB data is well documented in the literature and standard MCMC tools are publicly available, it is also a good reference point for the performance of the algorithm by Goodman and Weare as compared to a Metropolis-Hastings sampler.

\ch has been designed for optimal computational performance on large scale computing environments such as the Amazon  elastic compute cloud (EC2)\footnote{\url{http://aws.amazon.com/ec2/}}. We carry out a careful analysis of \ch's sampling efficiency, focusing in particular on the implications arising from the simultaneous sampling on multiple computers.

The architecture of our code makes \ch easily extendable to other applications, as it is straight forward to plug in Python modules containing further likelihood functions or codes for theory predictions.

We proceed as follows: In section \ref{sec:mcmcforcosmology}, the analysis of cosmological data using MCMC methods is briefly introduced and limitations of the current state-of-the-art are discussed. Section \ref{sec:emcee} explains how we address these limitations using the algorithm by \citet{Goodman2010a} and its implementation by \citet{Foreman-Mackey2012}. Introducing our code in section \ref{sec:implementation}, we review its components, explain its architecture and outline the parallelisation scheme. We test the code in section \ref{sec:testandresults} by sampling the WMAP 7 likelihood and comparing the results to MCMC chains from the Metropolis-Hastings sampler \cosmomc by \citet{Lewis2002}. In section \ref{sec:bench} we assess the performance of \ch on different cloud computing configurations. We discuss the results and conclude in section \ref{sec:conclusion}. Finally, we explain the installation of the package and give detailed examples for running the algorithm in the appendix.

%% file: problem.tex
\section{Markov chain Monte Carlo for cosmology}\label{sec:mcmcforcosmology}

In the following, we give a brief introduction to MCMC methods by discussing the Metropolis-Hastings (MH) algorithm as an example. Afterwards, we focus on the difficulties one faces when applying those methods to cosmological data.

\subsection{MCMC methods}

Assume we are interested in a probability density function $p(\theta)$ (called target distribution in the following) which is not given explicitly but can be calculated numerically up to a constant factor. If we want to learn about $p(\theta)$ we need to estimate it from a finite number of numerical evaluations. MCMC algorithms generate a sample distributed according to the target distribution in a probabilistic fashion. We illustrate it using the MH algorithm as an example. A classic textbook for further information is \citet{MacKay2003}, which is also on the web\footnote{\url{http://www.inference.phy.cam.ac.uk/itprnn/book.html}}.

Imagine the sampling process as a random walk in the parameter space of the target distribution. The walk has to be initialised at a point $\theta_0$ and a common method is to start at a random position close to the region where the likelihood is expected to be centered. In order to determine the next position, we need to specify a proposal density $P$ which has to be easy to sample and is usually chosen to be a Gaussian distribution. The probability of randomly proposing the new position $\theta'$ when being at $\theta_t$---the $t^{\text{th}}$ position in the sample---is given by $P(\theta'; \theta_t)$. After sampling $\theta'$ from $P$, the new position is accepted with probability
\begin{equation*}
	\min\left(1, \frac{p(\theta')}{p(\theta_t)} \frac {P(\theta_t; \theta')} {P(\theta'; \theta_t)} \right).
\end{equation*}
The updated $\theta_{t+1}$ is finally given by $\theta'$ if the step is accepted or by $\theta_t$ if it is rejected. The set $\{\theta_t\}_{t\in \{0, \cdots, T\}}$ converges to a sample from the target distribution $p(\theta)$ for large $T$ \citep{Metropolis1953, Hastings1970}.

It is worth noting that the efficiency of the sampling crucially depends on the chosen proposal distribution. If $P$ mainly proposes positions in the relevant parts of parameter space where the target distribution is large, the chain very quickly converges to a sample of $p(\theta)$. Yet, if the proposal distribution tends towards positions where $p(\theta)$ has low probability, most of the steps are rejected and samples will be heavily correlated. Using a proposal which is close to $p(\theta)$ is a convenient choice, as it guarantees that the proposed positions and the target are distributed similarly.

\subsection{Application to cosmology}

In cosmology the target distribution often arises from Bayesian inference. We consider the example of parameter estimation from CMB data for illustration. An observation measuring the CMB yields the angular power spectrum of the radiation as a result. At the same time, this power spectrum can be predicted from theoretical models for the evolution of the universe. The minimal concordance model, $\Lambda$CDM, depends on six parameters and it takes a few seconds to calculate it using a numerical Boltzmann integrator like \camb.

The likelihood function $\mathcal L$ of the parameters in the model arises from the conditional probability of the measurement result $X$ given the parameters $\theta = (\theta_1, \cdots, \theta_d)$:
\begin{equation*}
	\mathcal L(\theta) := p(X|\theta).
\end{equation*}
If we already have information on the parameters in the form of a prior distribution $q(\theta)$ from previous observations, $\mathcal L$ is used to update it according to Bayes' rule:
\begin{equation*}
	p^{new}(\theta) \propto \mathcal L (\theta)q(\theta).
\end{equation*}

Both $\mathcal L$ and the posterior $p^{new}$ depend on the results of the numerical Boltzmann integrator and are hence not available as analytical functions of $\theta$. As the number of parameters in this problem is usually greater than or equal to six, MCMC methods have to be used for estimating $\mathcal L$ or $p^{new}$. Although MCMC sampling makes the estimation feasible for large dimensions, it still suffers from the fact that calling the likelihood function---i.e. running the Boltzmann integrator for some specified parameters and comparing the results to the data in the CMB example---is costly in terms of time and resources. Consequently, we want our sampler to be as efficient as possible in terms of likelihood function calls per converged sample.

Additionally, MH sampling is by definition a serial process, iterating the calls to the likelihood function several thousand times. Using a single MH chain for inferring the parameters of $\Lambda$CDM from CMB data with an existing MCMC software package like CosmoMC, for example, takes at least a few hours on a notebook. Such runtimes are practical	in cases where a single parameter estimation run suffices, but quickly become limiting in general applications.

Testing how well various other models different from $\Lambda$CDM fit the data implies that the likelihood has to be explored separately for every distinct model. Furthermore, it is often necessary to introduce so-called nuisance parameters which model the process of data generation. Altering the number and the effect of the nuisance parameters is a way to test for the systematics in a measurement and requires the MCMC sampling process to be repeated multiple times. Whenever the MCMC analyses have to be iterated, it is hence desirable to decrease their run time---either by improving on the underlying MCMC process or by running the calculations in parallel on a cluster or cloud computing environment. In the next section we explain why the Python sampler \emcee is a good choice for achieving such a speed-up.

%% file: solution.tex
\section{Affine invariant ensemble sampler}\label{sec:emcee}

We learned in section \ref{sec:mcmcforcosmology} that MCMC sampling can be very time consuming in cosmological applications when the likelihood function is hard to evaluate numerically. Here we study how this can be overcome when using more recent MCMC algorithms such as the one by \citet{Goodman2010a} (called GW in the following) instead of MH. We therefore introduce the algorithm and its implementation by \citet{Foreman-Mackey2012} before discussing its advantages in terms of efficiency and parallelisation.

\subsection{The ensemble sampler by Goodman and Weare}

Instead of a single position which is updated during the course of the sampling, GW uses an ensemble of \emph{walkers} which are spread on the parameter space of the target distribution. At every iteration, the walkers are randomly assigned to a partner walker chosen from the ensemble and a random point on a line connecting their positions is proposed as the next step.

More formally, let $\theta^i_t$ denote the position of walker $i$ after $t$ iterations. When updating this position to $\theta_{t + 1}^i$, we pick another walker $\theta^j_t$ at random with $j \neq i$, sample a value $z$ from the \emph{fixed} distribution
\begin{equation*}
	q(z) = 	\left\{\begin{aligned}
				\frac 1 {\sqrt z} &\text{ if } z \in \left[ \frac 1 a, a \right]\\
				0 &\text{ otherwise}
			\end{aligned}\right.,
\end{equation*}
with tuning parameter $a$ and propose the position $\theta' = \theta^j_t + z(\theta^k_t - \theta^j_t)$. After evaluating the target distribution $p$ at the proposed position, we accept the step if
\begin{equation*}
	z^{d - 1}\frac {p(\theta')} {p(\theta^i_t)} \geq r,
\end{equation*}
with $r$ being a random number from $[0, 1]$ and $d$ the dimension of $p$.

GW is affine invariant, i.e. invariant under linear transformations of the target distribution. This implies in particular that the sampler is not sensitive to the scales of the parameters and does not depend on the covariances of the target distribution.

\subsection{Emcee}

The algorithm by \citet{Goodman2010a} was slightly altered and implemented as the Python module \emcee by \citet{Foreman-Mackey2012}. In this implementation, the algorithm does not update the walkers serially but instead divides them into two subgroups and updates all of the walkers from one subgroup at a time, using the other half of the walkers as their references.

\subsection{Advantages}\label{sec:emceeadvantages}

In section \ref{sec:mcmcforcosmology} we mentioned that MH needs a proposal distribution $P$ which is governing the efficiency of the algorithm. Thinking of $P$ as a $d$-dimensional Gaussian distribution with unknown covariance matrix, one finds that the MH has $d(d+1)/2$ tuning parameters. When the scales of variances and covariances of the target distribution are well known before the sampling, the tuning parameters are very helpful: By supplying the MH with an appropriate covariance matrix, the parameter space is explored efficiently and the sample quickly converges. Yet, if the target distribution is unknown and the covariance matrix is only a guess, the efficiency of the algorithm generally decreases. In case of non-Gaussian target distributions even a tuned Gaussian proposal might not match the target and efficiency can be low.

On the other hand, GW is affine invariant and thus not sensitive to the scales of the target distribution. Hence, the efficiency of the sampling process does not depend on having a good estimate of the target distribution before the sampling and a tuning of the algorithm is not necessary.

Equally important is the use of an ensemble of walkers instead of a single chain. As the \emcee implementation updates a large number of positions simultaneously at every iteration of the process, it can be parallelised in order to take advantage of a compute cluster (see section \ref{sec:implementation}).

The \emcee sampler based on GW thus has the potential to improve upon the limitations outlined in section \ref{sec:mcmcforcosmology}. First, we expect the sampling process to be fairly robust to changes in data and model. Second, it is possible to distribute the sampling on the multiple nodes of modern cluster or cloud computing environments. Both advantages can be achieved without tuning of the sampler or parallelisation of the underlying likelihood and theory codes.

%% file: components.tex
\section{Implementation of CosmoHammer}\label{sec:implementation}

We developed a Python framework called \ch for the estimation of cosmological parameters. The software embeds the Python package \emcee by \citet{Foreman-Mackey2012} and gives the user the possibility to plug in modules for the computation of any desired likelihood. The major goal of the software is to reduce the complexity when one wants to extend or replace the existing computation by modules which fit the user's needs as well as to provide the possibility to easily use large scale computing environments.

\subsection{Architecture}

\begin{figure*}[t]
\begin{center}

  \includegraphics[width=0.9\linewidth]{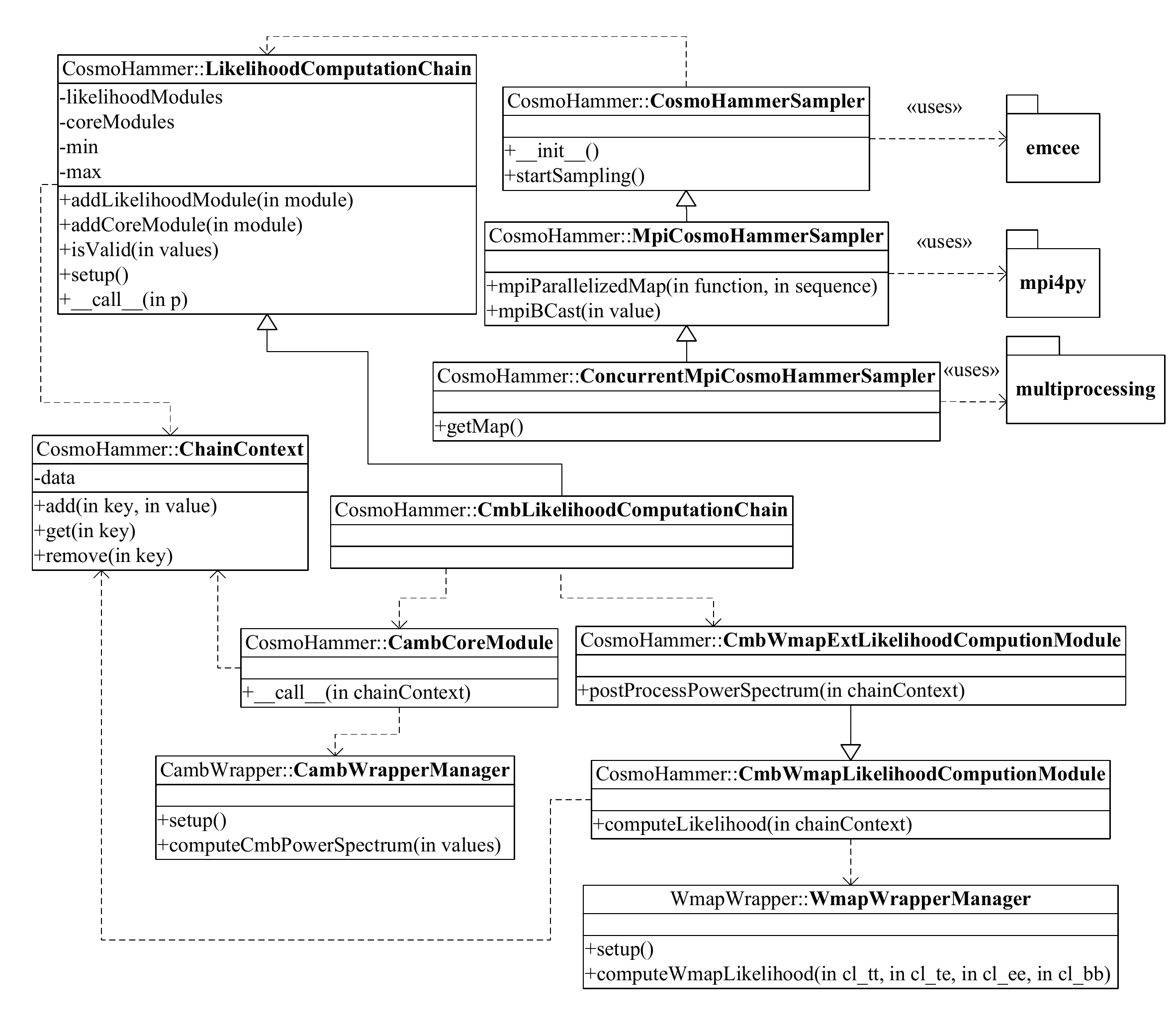}
  \caption{Class diagram of \ch's main architecture components.}
  \label{fig:classdiagram}
\end{center}
\end{figure*}

We applied the principle of chaining modules for the computation of the likelihood as depicted in Figure \ref{fig:classdiagram}. The class diagram shows the underlying architecture and the most important classes. The architecture allows for the development of self-contained and tested modules which can be assembled in different sampling configurations. The internal design separates parameter space exploration, theory prediction and likelihood computation. This makes it easy to extend or replace these modules by new algorithms.

The concept of \ch divides the modules in two logical groups: modules for the computation of the likelihood and core modules. The core modules produce information like the CMB power spectrum or other theory predictions for the likelihood modules. The individual core modules can be combined in an instance of the \emph{LikelihoodComputationChain} module (see Figure \ref{fig:classdiagram}). The chain stores the modules and initialises them in the right order. Furthermore, it ensures that the sampled parameters stay within physically motivated bounds during the sampling process.

The modules in the \emph{LikelihoodComputationChain} communicate via the \emph{ChainContext} in which arbitrary data can be stored and retrieved. This minimises the dependencies between the individual modules and ensures that they can be replaced without the need to change or extend \ch.

\subsection{Components}
\ch comes with a set of modules which compute the CMB power spectrum and the WMAP likelihood by wrapping the theory prediction code \camb by \cite{Lewis2000} and the likelihood code from the WMAP team \citep{Larson2011, Komatsu2011}. It integrates the two Fortran modules by wrapping them using numpy's F2PY\footnote{\url{http://www.scipy.org/F2py}}. The Fortran to Python interface generator provides the connection between Python and Fortran. In this way we can take full advantage of the well tested and widely used Boltzmann integrator \camb as well as of the WMAP likelihood computation module while combining them with the \emcee sampling algorithm. Furthermore, by using the wrapped code we benefit from the performance of Fortran in the convenient Python environment.

\subsubsection{CAMB and WMAP modules}
To use \camb and the WMAP code we create an instance of the \emph{LikelihoodComputationChain} and add an instance of the \emph{CambCoreModule} and \emph{CmbWmapLikelihoodComputationModule}. The \emph{CambCoreModule} delegates the computation of the power spectrum to the \emph{CambWrapperManager} and the \emph{CmbWmapLikelihoodComputationModule} in turn delegates the likelihood computation to the \emph{WmapWrapperManager}.

Alternatively an instance of the preconfigured \emph{CmbLikelihoodComputationChain} can be created. This chain extends the regular \emph{LikelihoodComputationChain} and ensures the correct setup of the two modules. A detailed example is given in \ref{sec:examples}.

\subsubsection{Samplers}
\ch also provides different samplers. The samplers embed the \emcee package and are responsible for logging and storing the obtained results. The \emph{CosmoHammerSampler} is used when running on a single physical computer using one or multiple threads. The functionality, however, is limited to a single computer. The extended \emph{MpiCosmoHammerSampler} provides the required functionality when \ch should take advantage of a computation cluster with multiple physical nodes like a cloud or grid computer. This sampler uses Message Passing Interface (MPI)\footnote{\url{http://mpi4py.scipy.org/}} for the communication between the nodes in the cluster.
 
The implementation uses the paradigm of workload partitioning in which the work is split into blocks of nearly equal length. Every node then processes its block and returns the results to the master node. The master node gathers all results and merges them into a single list which then is returned to \emcee.
 
When using a compute cluster the nodes often come with a large number of computational cores. Writing code that fully benefits from such a large number of cores is usually difficult. Therefore, it makes sense to split the workload also on the node since using a smaller number of cores per computation while performing multiple computations in parallel is typically more efficient. In section \ref{sec:benchmarks} it can be seen how the execution time decreases when \ch uses multiple processes on one physical node.
 
If it is desired to distribute the workload to several nodes in a cluster as well as to spawn multiple processes on a node, the provided \emph{ConcurrentMpiCosmoHammerSampler} can be used. This sampler introduces another level of parallelisation by using Python's built in multiprocessing package.

\subsubsection{Initialisation}

By default all samplers initialise their walkers in a ball around a given center point as suggested in \cite{Foreman-Mackey2012}. The starting position $\theta^i_0$ of every walker $i$ is then computed as follows:
\begin{equation*}
	\theta^i_0 = P_c+N(0,1)*P_w,
\end{equation*}
where $P_c$ is the center point and $P_w$ is the start width. Both $P_c$ and $P_w$ have to be supplied by the user for every parameter. Furthermore, \ch comes with a built in generator producing a top-hat distribution as follows:
\begin{equation*}
	\theta^i_0 = P_c + \epsilon * P_w,
\end{equation*}
where $\epsilon$ is a pseudorandom number ranging from -1 to 1.

If one prefers to launch the sampling process with a different starting strategy, it is possible to pass a custom implementation of a \emph{PositionGenerator} to the \emph{CosmoHammerSampler} instance.

%% file: testsandresults.tex
\section{Tests and results}\label{sec:testandresults}

For testing the efficiency of \ch we compare it to \cosmomc, a widely used MCMC engine for cosmological parameter estimation by \citet{Lewis2002}. \cosmomc is a Fortran90 code which also uses \camb and the WMAP likelihood code for estimating cosmological parameters from CMB data, but it employs the MH algorithm for creating its MCMC chains. It furthermore contains an extensive choice of additional datasets and analysis modules.

To optimise the sampling process, \citet{Lewis2002} use the following seven default parameters for describing the $\Lambda$CDM model: the physical baryon density $\Omega_b h^2$, the physical dark matter density $\Omega_{DM} h^2$, the ratio of the approximate sound horizon to the angular diameter distance $\theta$, the reionisation optical depth $\tau_{re}$, the scalar spectral index $n_s$, the primordial superhorizon power in the curvature perturbation on $0.05Mpc^{-1}$ scales $\Delta_R^2$, and finally $A_{sz}$, a Sunyaev-Zel'dovich template normalisation. Furthermore, some of the parameters are rescaled to end up with similar orders of magnitude.

As \emcee is affine invariant, the particular choice of parametrisation is not expected to influence its performance. For simplicity, we decided to use the same ones as \cosmomc up to rescaling, yet replacing $\theta$ by Hubble's constant $H_0$.

\subsection{Analysis of MCMC chains}\label{sec:analysis}

Following the lines of \cite{Foreman-Mackey2012} we adopt the \emph{autocorrelation time} as our primary quality criterion for an MCMC sampler. Consider a probability density function $p(\theta)$, an MCMC sample $\{\theta_t\}$ of this distribution, and a function $f(\theta)$ whose mean $\langle f \rangle = \int f(\theta) p(\theta)\, d\theta$ we wish to estimate. The autocorrelation $C_{ff}$ of $f(\theta)$ evaluated at the sample points $\{\theta_t\}$ with lag $T$ is defined as:
\begin{equation*}
	C_{ff}(T) := \langle (f(\theta_t) - \langle f \rangle)(f(\theta_{t + T}) - \langle f \rangle) \rangle.
\end{equation*}
Typically, the autocorrelation of an MCMC sample is non-zero and decaying with increasing lag: $C_{ff}(T) \propto \exp(-\frac T {\tau_{ff}})$. However, if the points $\{\theta_t\}$ were independent of each other, the autocorrelation function would vanish for all $T \geq 1$.

Let us also define the normalised autocorrelation function
\begin{equation*}
	\rho_{ff}(T) = \frac {C_{ff}(T)} {C_{ff}(0)},
\end{equation*}
where $C_{ff}(0)$ is the variance of the sample $\{f_t\} := \{f(\theta_t)\}$.
There are two relevant timescales connected to $\rho_{ff}(T)$. On the one hand, there is the exponential autocorrelation time $\tau_{exp}$ which is defined by
\begin{equation}
	\tau_{exp} = \underset{T \rightarrow \infty}{\lim\sup} \frac T {-\log\left|\rho_{ff}(T)\right|}.
	\label{eq:tauexp}
\end{equation}
On the other hand, the integrated autocorrelation time $\tau_{int}$ is given by
\begin{equation}
	\tau_{int} = \frac 1 2 + \sum_{T = 1}^{\infty}\rho_{ff}(T).
	\label{eq:tauint}
\end{equation}
In general $\tau_{exp}$ and $\tau_{int}$ are not equivalent, although they are both equal to $\tau_{ff}$ for exponentially decaying autocorrelation functions $C_{ff}$.

The reason why $\tau_{exp}$ and $\tau_{int}$ are relevant for the analysis of MCMC samples is connected to the issues of thermalisation at the beginning of an MCMC chain (often called \emph{burn-in}) and the statistical errors one has to account for when evaluating the expectation value of $f(\theta)$ using the sample $\{\theta_t\}$ (see the lecture notes by \citet{Sokal1989} for more information).

The exponential autocorrelation time tells us how many iterations should be discarded at the beginning of the Markov chain in order to avoid an initialisation bias, since it measures the time we have to wait for two positions in the sample $\{f_t\}$ to be close to uncorrelated. Discarding a few exponential autocorrelation times at the beginning of the sampling typically suffices to suppress the bias.

At the same time, the integrated autocorrelation time determines the standard error of the mean through:
\begin{equation}\label{eq:errorinthemean}
	\var(\bar f) = \frac {2\tau_{int}} {N} \var(f_t),
\end{equation}
where $N$ is the size, $\bar f$ is the mean, and $\var(f_t)$ is the variance of the sample $\{f_t\}$.

An estimate for $C_{ff}(T)$ is given by
\begin{equation}\label{eq:cest}
	C_{ff}(T) \approx \hat C_{ff}(T) = \frac 1 {N - T} \sum_{t = 1}^{N - T}(f_t - \bar f)(f_{t + T} - \bar f).
\end{equation}
Estimating the integrated autocorrelation time from \eqref{eq:cest} is not easy because of issues regarding the statistical noise in the large $T$ limit of $\hat C_{ff}(T)$. Yet, the autocorrelation function for our problem is exponentially decaying, meaning that we can also evaluate $\tau_{exp}$ instead of $\tau_{int}$. We find that estimating $\tau_{int}$ from a fit to $\hat C_{ff}(T)$ is the best choice for evaluating the autocorrelation time of this particular problem (see \ref{sec:estimation_of_autocorrelation} for more information). As $\tau_{exp}$ and $\tau_{int}$ are equivalent for our purposes, we denote both as the autocorrelation time $\tau$ for simplicity.

\subsection{Multiple walkers}\label{sec:multiple}

We typically demand that MCMC samples satisfy the following accuracy criterion: The statistical error of the mean $\bar f$ we wish to estimate---defined in equation \eqref{eq:errorinthemean}---has to be smaller than a given fraction $\epsilon$ of the standard deviation of its marginal distribution. In other words
\begin{equation}
	\sqrt{\frac {\var (\bar f)} {\var (f)}} = \sqrt{\frac {2\tau} {N}} \leq \epsilon
	\label{eq:stoppingcrit}
\end{equation}
and consequently
\begin{equation}
	N \geq \frac {2\tau} {\epsilon^2},
	\label{eq:ncond}
\end{equation}
where $N$ is the total size of the sample. For multiple walkers or independent chains, $N$ is given by the number of steps per walker or chain $n$ (called sequential steps in the following) times the number of walkers $L$ and hence
\begin{equation}
	n \geq \frac {2\tau} {\epsilon^2 L}.
	\label{eq:ncond2}
\end{equation}

Equation \eqref{eq:stoppingcrit} only holds when the sample is unbiased by the initialisation and this is true in the asymptotic limit of large $n$. When sampling on a cloud computing infrastructure, however, we need a large number of walkers $L$ for maximum parallelisability and hence expect rather small $n$ according to \eqref{eq:ncond2}.

Yet, the sample gets close to unbiased when discarding the initial steps of every walker or chain as burn-in. We already mentioned in section \ref{sec:analysis} that such a burn-in phase is expected to last for a few autocorrelation times. When using only a few walkers, i.e. $L$ close to one, $n$ is large and the burn-in phase is a subdominant part of the overall sampling. As $L$ grows, though, the number of sequential steps $n$ becomes comparable to the burn-in length and the discarded samples turn into a dominant fraction of the overall sample size. We analyse the consequences of this result in section \ref{sec:efficiency}.

\subsection{Configuration of CosmoMC and CosmoHammer}\label{sec:config}

In the following we want to compare \ch to two different configurations of \cosmomc. In each case we sample the likelihood given by the Fortran90 code and the data of the WMAP 7 team.

\begin{table}[t]
	\centering
	\caption{Each \emcee walker and \cosmomc chain is initialised at a random position around the mean, within a deviation sampled from a Gaussian distribution with given width.}
	\label{tab:startDistribution}
	\resizebox{\linewidth}{!}{
		\begin{tabular}{l|c|c|c|c|c|c|c|c}\hline
			Parameter & $H_0$ & $\Omega_{b} h^2$ & $\Omega_{DM} h^2$ & $10^{9}\Delta_R^2$ & $n_s$ & $\tau_{re}$ & $A_{SZ}$ & $\theta$ \\ \hline
			Mean & 70 & 0.0226 & 0.122 & 2.1 & 0.96 & 0.09 & 1 & 1.04 \\ \hline
			Width & 3 & 0.001 & 0.01 & 0.1 & 0.02 & 0.03 & 0.4 & 0.002 \\ \hline
			Lower bound & 40 & 0.005 & 0.01 & 1.48 & 0.5 & 0.01 & 0 & .5 \\ \hline
			Upper bound & 100 & 0.1 & 0.99 & 5.45 & 1.5 & 0.8 & 2 & 10 \\ \hline
		\end{tabular}
	}
\end{table}

The first \cosmomc instance is an out-of-the-box approach, using the standard configuration shipped with the \cosmomc package. It initialises the chain in a small ball around an estimated mean of the likelihood using the numbers from Table \ref{tab:startDistribution} and supplies a covariance matrix to specify the Gaussian distribution which is used as the proposal. We will refer to this configuration as \emph{fine-tuned \cosmomc}.

The second approach employs an option of \cosmomc which splits the sampling into two phases. Starting with an initial guess for the covariance matrix in the tuning phase---a diagonal matrix with estimated variances in our case---the sampler continuously updates the proposal's covariance matrix from the last half of the generated samples. Afterwards, \cosmomc uses the generated proposal to create the samples for estimating the likelihood. This approach will be called \emph{self-tuning \cosmomc} from now on. When using multiple independent chains, \cosmomc allows one to automatically stop the tuning phase once a convergence criterion is fulfilled. We use this option to ensure that the sampling after the tuning phase is comparable to the fine-tuned \cosmomc process. For our analyses, we used 10 independent chains for both \cosmomc configurations.

For \ch we call \camb in exactly the same fashion as the standard \cosmomc configuration does (i.e. we use the same theoretical model for our cosmology) and also initialise the walkers according to Table \ref{tab:startDistribution}. We furthermore used \emcee with 350 walkers, a rather arbitrary pick which was convenient to work with.

The bounds listed in Table \ref{tab:startDistribution} are needed to ensure that the parameters which are passed to the Boltzmann integrator \camb make sense physically. Whenever the sampler proposes a position which is out of bounds, \ch returns zero probability immediately.

\subsection{Efficiency}\label{sec:efficiency}

\begin{table}[t]
	\centering
	\caption{Estimated autocorrelation time $\tau$ for the seven parameters and the three sampler configurations.}
	\label{tab:tauexp}
	\resizebox{\linewidth}{!}{
		\begin{tabular}{c|c|c|c|c|c|c|c}\hline
			Sampler & $H_0$ & $\Omega_b h^2$ & $\Omega_{DM} h^2$ & $\Delta_R^2$ & $n_s$ & $\tau_{re}$ & $A_{SZ}$\\ \hline
			\ch & 44.4 & 44.7 & 44.4 & 43.4 & 43.8 & 43.8 & 48.2\\ \hline
			fine-tuned \cosmomc & 17.3 & 17.1 & 15.1 & 13.8 & 17.6 & 14.4 & 18.4 \\ \hline
			self-tuning \cosmomc & 16.6 & 14.9 & 19.1 & 13.3 & 15.0 & 14.8 & 16.3 \\ \hline
		\end{tabular}
	}
\end{table}

A good estimate for the autocorrelation function is important for the analysis of the sample. For both \cosmomc and \ch the autocorrelations
\begin{equation*}
	\hat C_{X}(T) = \frac 1 {n - T} \sum_{t = 1}^{n - T}(X_t - \bar X)(X_{t + T} - \bar X),
\end{equation*}
with $X$ being one of the seven dimensions of the parameter-space, behave equivalently for all parameters. As the estimation of the autocorrelation times for the different sampler configurations is not straightforward, we give a detailed description of our procedure in \ref{sec:estimation_of_autocorrelation}. The results can be found in Table \ref{tab:tauexp}. It is not very surprising that the \cosmomc processes are more efficient in terms of calls per independent sample, as they are using a well tuned proposal to sample a target distribution which is itself close to normally distributed. On the other hand, \emcee needed no tuning while still performing reasonably well in terms of autocorrelation time.

\begin{figure}[t]
	\centering
	\includegraphics[width=.9\linewidth]{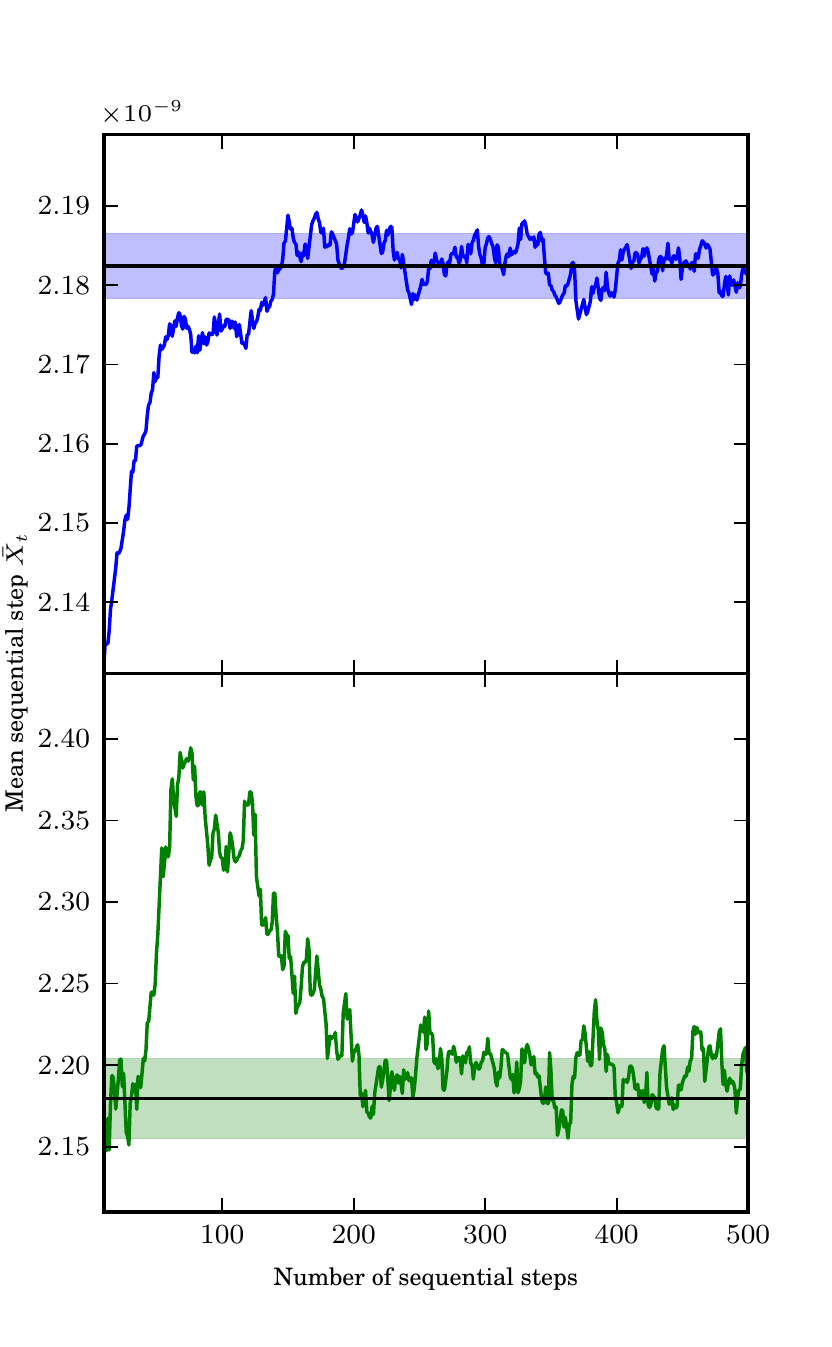}
	\caption{Mean sequential steps $\bar X_t$ of \ch (top) and fine-tuned \cosmomc (bottom) for parameter $\Delta_R^2$. The shaded regions depict the error in the mean $\sigma(\bar X_t)$ for every mean sequential step. Self-tuning \cosmomc runs need about 2000 initial steps for tuning, but start at an unbiased position afterwards.}
	\label{fig:burnin}
\end{figure}

Additionally we need to consider the optimal length of the burn-in period before the actual sampling starts. As was discussed in section \ref{sec:multiple}, this is particularly important if we want to iterate a large number of walkers. Let $\{X_t^i\}$ denote the position of walker $i\in\{1, \cdots, L\}$ at iteration $t\in\{1, \cdots, n\}$ for parameter $X$. For finding the burn-in length, we observed the mean sequential step $\bar X_t$
\begin{equation*}
	\bar X_t = \sum_{i = 1}^L X_t^i
\end{equation*}
for increasing time $t$. Once the positions of the walkers are drawn from the target distribution in an unbiased way, we expect $\bar X_t$ to vary around the true mean as predicted by the standard error in the mean
\begin{equation*}
	\sigma(\bar X_t) = \frac \sigma {\sqrt{L}},
\end{equation*}
where $\sigma$ is the standard deviation of the marginal target distribution for parameter $X$. As long as this is not the case, the walkers are still biased by the initialisation.

Using this method we find that a safe choice for the burn-in period is 250 for both the fine-tuned \cosmomc and \ch, as can be seen in Figure \ref{fig:burnin} for parameter $\Delta_R^2$. This is surprising as both processes have very different autocorrelation times, but the MCMC algorithm of \emcee turns out to be very efficient in getting rid of its initialisation bias. The self-tuning algorithm needs about 2000 iterations to estimate its proposal distribution, but already starts at an unbiased position afterwards. The reason for such a long tuning phase is that the sampling efficiency is very poor when the sampler is untuned in the beginning.

\begin{figure*}[t]
	\centering
	\includegraphics[width=1.\linewidth]{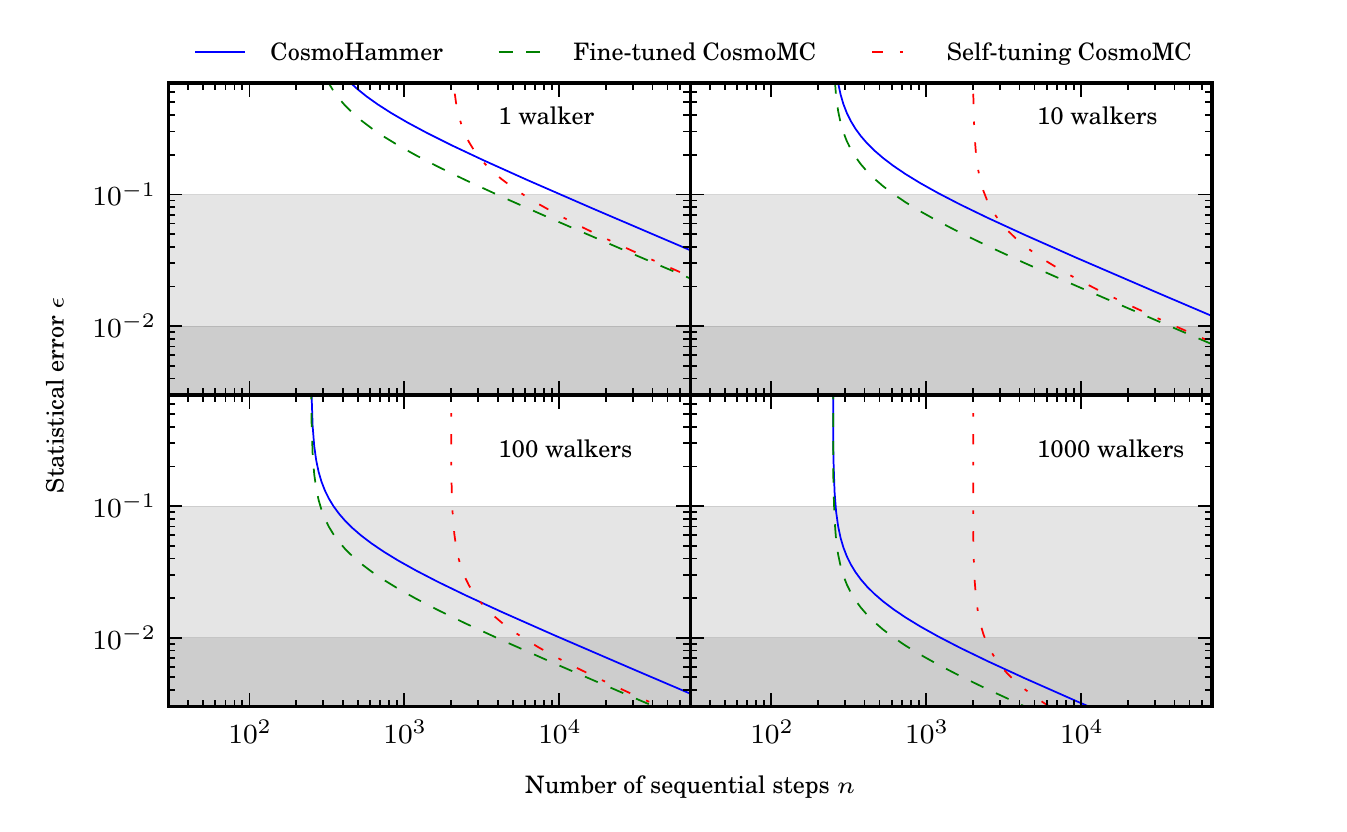}
	\caption{Statistical error $\epsilon$, defined in equation \eqref{eq:epsilon}, as a function of sequential steps $n$ and for different numbers of walkers. For simplicity, we assume in this plot that burn-in length and autocorrelation time are independent of the number of walkers or chains. The shaded regions show where the respective configurations reach statistical error $\epsilon = 10\%$ and $\epsilon = 1\%$.}
	\label{fig:stater}
\end{figure*}

We can now consider the statistical error $\epsilon$ from equation \eqref{eq:ncond2} as a function of burn-in or tuning length $b$, sequential steps $n$, number of walkers $L$, and autocorrelation time $\tau$:
\begin{equation}
 	\epsilon = \sqrt{ \frac {2\tau} {(n - b)L} } \text{ for } n > b.
 	\label{eq:epsilon}
 \end{equation}
This is visualised in Figure \ref{fig:stater} for all sampler configurations, using $\tau$ from table \ref{tab:tauexp} and the corresponding burn-in and tuning values.

We find that the most efficient way to create a sample which satisfies the accuracy criterion \eqref{eq:stoppingcrit} is to use a fine-tuned \cosmomc. This is not surprising, as it has a good estimate of the target distribution before it even starts the sampling process, resulting in a short autocorrelation time and burn-in phase. Yet, a well tuned MH is typically not available when analysing new data. In this case, the self-tuning \cosmomc configuration or similar concepts with lengthy tuning phases have to be used for configuring the sampler.

It is this tuning phase which turns out to be the bottle-neck for the parallelisation of a MH sampler. We can see in Figure \ref{fig:stater} that for as few as 10 walkers, \ch reaches the 10\% error regime $\epsilon \leq 0.1$ before the self-tuning \cosmomc run even finalises the tuning at $n = 2000$. When estimating the parameter $X$ from the target distribution, the result usually reads
\begin{equation*}
	\hat X = \bar X \pm \sigma(X),
\end{equation*}
with estimate $\hat X$, sample mean $\bar X$, and standard deviation $\sigma(X)$. Consequently, the statistical error does not affect $\hat X$ up to the second digit of $\sigma(X)$ when $\epsilon$ is smaller than $0.1$ and is hence sufficiently small for most applications in cosmological parameter estimation.

\subsection{Parameter estimation}\label{sec:paramest}

We know from the previous discussion that the errors we expect for our estimates of the mean are of the order $\sigma\sqrt{2\tau / N}$, where $\sigma$ is the standard deviation of the target distribution for the respective parameter. The total sample size after the burn-in was chosen to be $N = 250 \times 350 = 87,500$ for \ch, predicting an accuracy in the mean estimate of about $3.4\%$ relative to the standard deviation. The same precision will be reached when using about $N = 3,500 \times 10 = 35,000$ samples from \cosmomc.

\begin{table}[t]
	\centering
	\caption{Mean, standard deviation and the expected statistical error of the parameters sampled by \ch and \cosmomc.}
	\label{tab:sampledResults}
	\resizebox{\linewidth}{!}{
		\begin{tabular}{c|c|c|c}
			\hline
			Parameter & CosmoHammer & \cosmomc & Statistical error\\ \hline
			$H_0$ & $ 70.4^{+2.8}_{-2.5} $ & $ 70.2^{+2.4}_{-2.1} $ & $\pm 0.1$\\ \hline
			$100\Omega_{b0}h^2$ & $ 2.247^{+0.065}_{-0.051} $ & $2.243^{+0.057}_{-0.057} $ & $\pm 0.002$ \\ \hline 
			$\Omega_{dm0}h^2$ & $0.1107^{+0.0063}_{-0.0049} $ & $ 0.1116^{+0.0048}_{-0.0055} $ & $\pm 0.0002$ \\ \hline
			$10^{9}\Delta_R^2$ & $ 2.174^{+0.078}_{-0.069} $ & $2.168^{+0.077}_{-0.068} $ & $\pm 0.003$ \\ \hline 
			$n_s$ & $ 0.967^{+0.014}_{-0.014} $ & $0.966^{+0.014}_{-0.012} $ & $\pm 0.0005$ \\ \hline 
			$100\tau_{re}$ & $ 8.7^{+1.5}_{-1.3} $ & $8.6^{+1.5}_{-1.3} $ & $\pm 0.05$ \\ \hline 
			$A_{SZ}$ & $ 0.92^{+0.71}_{-0.64} $ & $ 1.01^{+0.65}_{-0.72} $ & $\pm 0.02$ \\ \hline
		\end{tabular}
	}
\end{table}

\begin{figure*}[p]
	\centering
	\includegraphics[width=.8\textwidth]{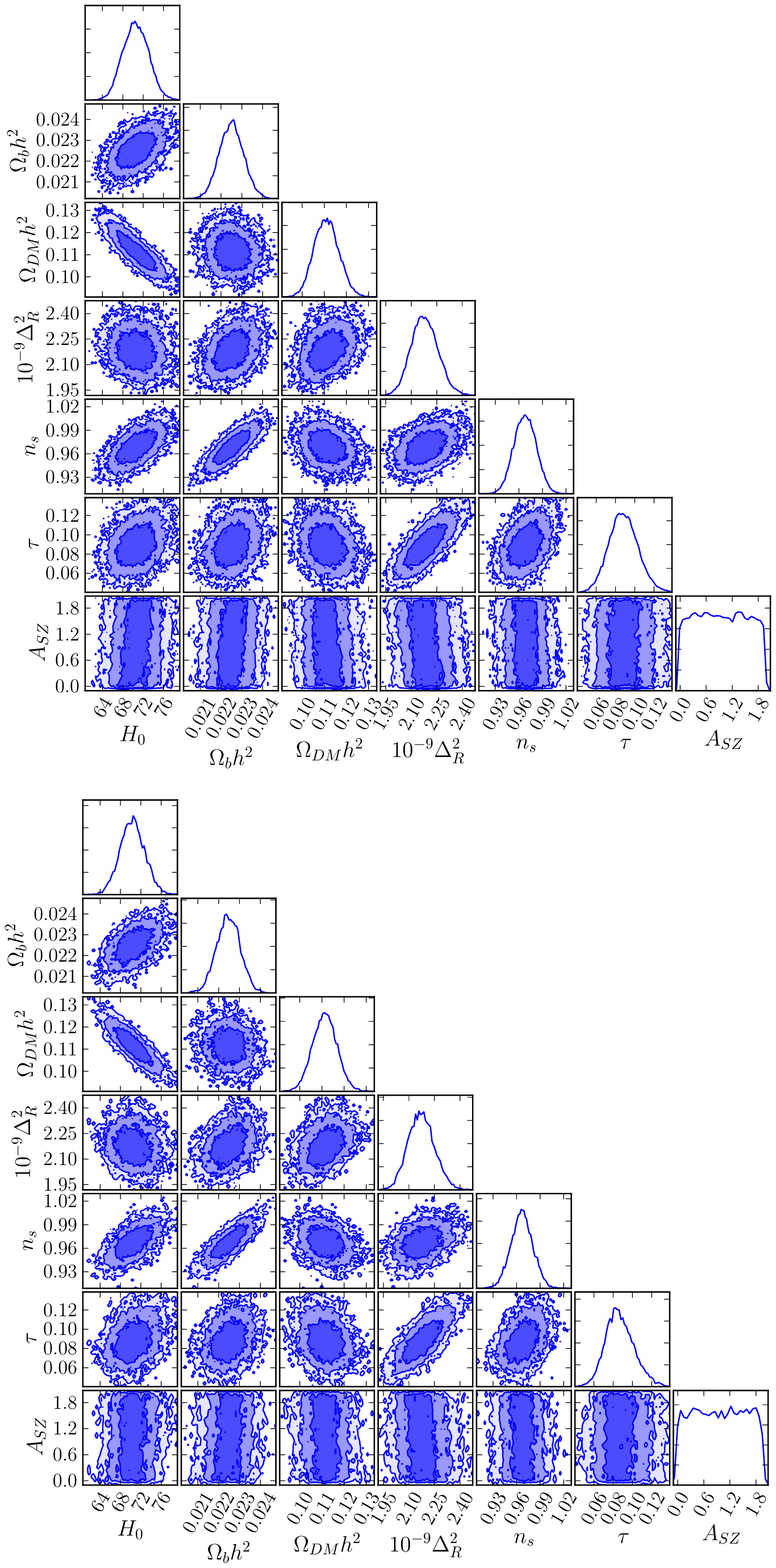}
	\caption{One and two dimensional marginal distributions of the WMAP 7 likelihood as sampled by \ch (top) and \cosmomc (bottom) for all parameter combinations.}
	\label{fig:distributions}
\end{figure*}

As we chose $N$ such that the error is smaller than 3.4\% of the standard deviation, the parameter estimates of the different sampler configurations are supposed to vary on this scale, too. We can see from Table \ref{tab:sampledResults} that this is indeed the case. Finally, Figure \ref{fig:distributions} shows the projections of the 7-dimensional likelihood into one and two dimensional marginal distributions.

We conclude that the samples of \cosmomc and \ch behave just as expected from our analysis in section \ref{sec:analysis}. In particular, this means that the quality of the sampling is well understood once the autocorrelation of the MCMC process is known. Tests based on multiple independent instances of the employed sampler configurations support these conclusions.

%% file: benchmarks.tex
\section{Performance measures and metrics}\label{sec:bench}

The performed computations required a large amount of computational power. We therefore decided to explore the possible benefits of cloud computing by means of \ch. One of the major advantages of this computing strategy is that the configuration of the cloud can be easily tailored to the problem at hand. In the cloud more computational power can be added within minutes by renting extra compute instances on demand, resulting in an optimised execution time. The following section describes the environment and the configuration used to perform the benchmarks.

\subsection{Environment}
As cloud service provider we decided to use Amazon EC2 in combination with the Starcluster Toolkit (Software Tools for Academics and Researchers)\footnote{\url{http://star.mit.edu/cluster/}}. Table \ref{tab:instancetypes} shows the configuration of the instance types we used for the benchmarks. The high performance computing cluster consisted of one master node and several worker nodes. At the moment of the benchmarks one cc2.8xlarge Instance ships with 2 $\times$ Intel Xeon E5-2670, eight-core architecture with Hyper-Threading, resulting in 32 cores per node. We used a m1.large instance as master node mainly to benefit from the high I/O performance in order to reduce the loading time of the WMAP data.

\begin{table}[t]
	\centering
	\caption{Amazon EC2 instance types used for the benchmarks in Figure \ref{fig:benchmark}.}
	\label{tab:instancetypes}
	\resizebox{\linewidth}{!}{
	\begin{tabular}{c|c|c}\hline
		 & Master node & Worker node \\ \hline
		Name & Large Instance & Cluster Compute Eight Extra Large\\ \hline
		API Name & m1.large & cc2.8xlarge \\ \hline
		Memory & 1.7 GB & 60.5 GB \\ \hline
		Instance storage & 160 GB  & 3370 GB  \\ \hline
		Processing Power & 2 Cores & 32 Cores \\ \hline
	\end{tabular}
	}
\end{table}

To compile the native Fortran modules, we used Intel's ifort compiler and mkl libraries, Python 2.7, and numpy 1.6.2. 
 
\subsection{Benchmarks}\label{sec:benchmarks}

\begin{figure*}[t]
\begin{center}
   \includegraphics[width=0.65\linewidth]{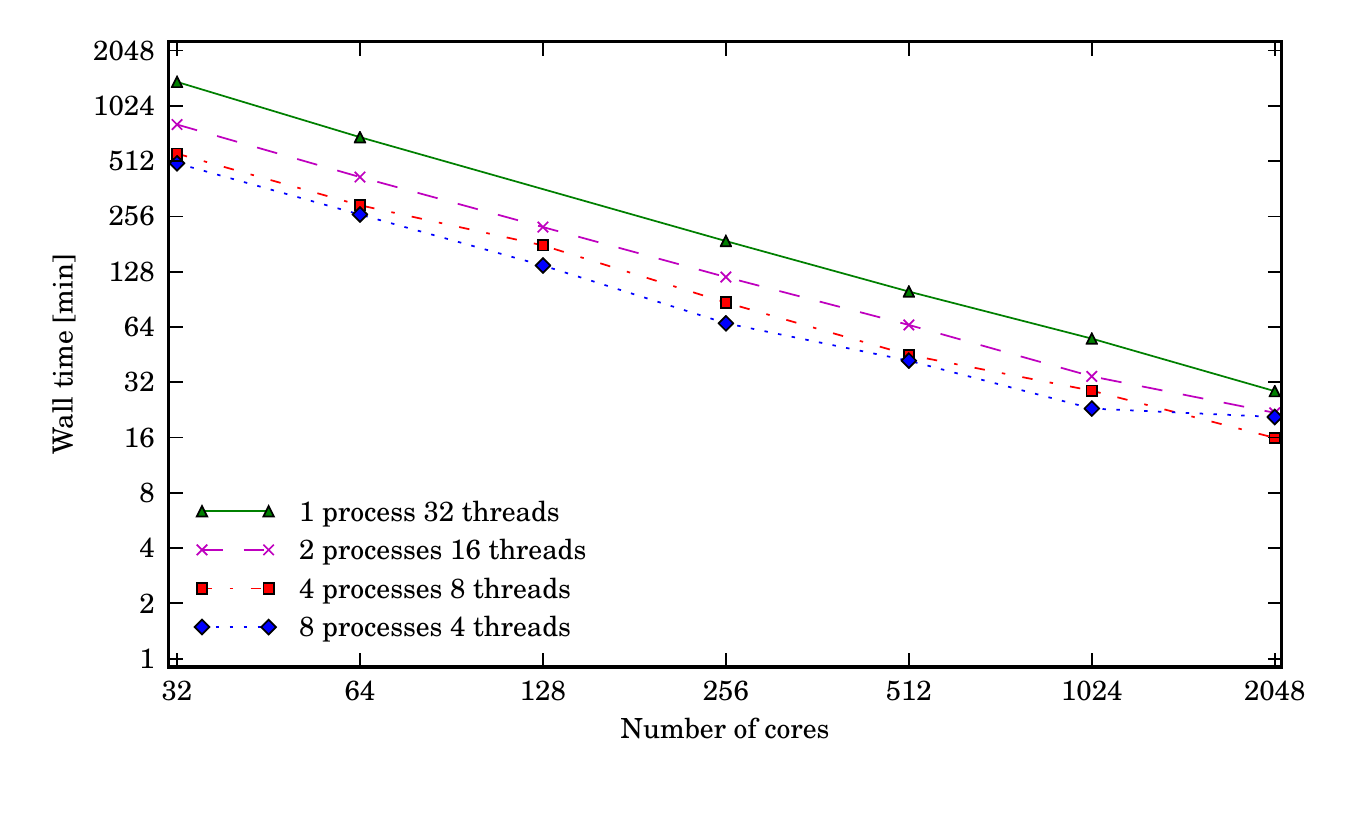}
   \caption{Run time behaviour of \ch with changing number of cores using different parallelisation schemes.}
   \label{fig:benchmark}
\end{center}
\end{figure*}

The results depicted in Figure \ref{fig:benchmark} have been realised with one to 64 worker nodes (32 - 2048 cores) and different combinations of processes and threads per node. The processes define the number of computations executed in parallel and the threads represent the number of cores used for one computation. In the case of 4 processes and 8 threads, for instance, there were four Python processes working in parallel on every node, each of them spawning eight threads.
 
For all test runs, \ch was configured to use 350 walkers and to run 500 sampling iterations, resulting in $175,000$ samples per run. This sample size was also used for the parameter estimation in section \ref{sec:paramest}. We used a \emph{LikelihoodComputationChain} with the \emph{CambCoreModule} and the \emph{CmbWmapExtLikelihoodComputationModule} for the computation of the likelihood.

Figure \ref{fig:benchmark} shows that the wall time $T$ as a function of number of cores $N$ behaves as a power law: $T \propto N^{-\alpha}$, i.e. it is linear on a logarithmic scale with a mean $\alpha$ of $0.89$. The best result was achieved using 64 nodes with 32 cores, four processes and eight threads. Using this configuration, the computation took about 16 Minutes.

\subsection{Discussion}\label{sec:discussion}

The implemented algorithm for the parallelisation is efficient yet easy to understand. Note, however, that this way of parallelisation is only beneficial when the executed computations are time and resource consuming. Distributing the workload in a compute cluster always implies the transfer of information over the network which is typically slower than transferring information between local processes by an order of magnitude. Therefore, the advantage of additional computing resources and the disadvantage of the network overhead have to be weighted.

Since we use \ch for the estimation of cosmological parameters which implies the execution of computationally intensive theory prediction modules like \camb, the overhead of the network latency plays a secondary role. We expect that \ch will be extended by additional theory and likelihood modules so that the network overhead can be neglected and the number of computational nodes and therefore CPU's can be increased. Nevertheless, there is a given maximum: The number of nodes cannot exceed the number of walkers.

\ch can not only take advantage of multiple nodes in a cluster but can be parallelised also on every node. As illustrated in Figure \ref{fig:benchmark}, there is a maximum for the number of processes on the node, too. The test run using eight processes and four OMP threads reaches a plateau when using 64 nodes. This is caused by the workload partitioning where the 350 walkers are equally distributed on 64 nodes. Consequently, more processes are available than walkers to process, which in turn causes the processes and their spawned threads to idle instead of being used for the computation of the power spectrum.

Inferring parameters from WMAP data takes 30 hours on a dual core notebook with a single MH process. Using one chain, 37,000 samples need to be created to obtain results in the specified error regime. When parallelising an MCMC process the scalability is affected by two overheads: First, the number of samples has to be increased to $N = 2 \times 250 \times 350 = 175,000$ due to the growth of burn-in overheads and a larger autocorrelation time as described in section \ref{sec:paramest}. This consequently results in additional computational costs. Second, the distribution of the workload in the compute cluster causes a loss of efficiency. In the benchmarks, a power law behaviour with an exponant of $\alpha = 0.89$ was measured, resulting in a scaling overhead of $11\%$ when the number of cores is doubled. Accounting for both effects, the wall time decreases to 16 minutes on 2048 cores.

%% file: conclusion.tex
\section{Conclusion}\label{sec:conclusion}

In cosmological applications, accelerating MCMC methods is crucial whenever the sampling has to be repeated numerous times and evaluating the target distribution is computationally expensive. Examining the measurement systematics of an astrophysical observation in an iterative analysis, for example, requires sampling an extensive amount of target distributions, each depending on time intensive simulations. Whenever it is not possible to massively parallelise the likelihood code, starting a large number of short chains enables us to evolve them in parallel not only on a couple of local CPUs, but globally on a large scale computing infrastructure.

We found that two characteristics are crucial for the performance of parallelised MCMC sampling: It has to be efficient in terms of autocorrelation time and---equally important---it has to sample efficiently without needing much overhead for tuning and equilibration. It is this overhead which puts a fundamental lower bound on the run time of the sampling procedure and can limit the application of MCMC methods when a large number of target distributions has to be explored. The \emcee sampler by \citet{Foreman-Mackey2012} turned out to be a good choice regarding those preconditions: It not only requires no tuning of the algorithm in general, but also avoids an initialisation bias very quickly while performing fairly well in terms of autocorrelation time when applied to the likelihood of the WMAP 7 observations.

In order to exploit the advantages of \emcee on large cloud computing configurations or similar infrastructure, we introduced \ch, a Python framework for parallelised MCMC sampling. It enabled us to explore arising computer science technologies like elastic cloud computing for scientific applications. The elastic nature of Amazon EC2, for example, ensures that scientists and engineers do not have to wait in long queues to access shared clusters as the computational power can be increased within minutes by renting additional compute instances. We reduced the time for estimating cosmological parameters using the WMAP 7 likelihood from 30 hours on a desktop computer running a MH sampler to about 16 minutes on a cluster with 2048 cores on Amazon EC2. Furthermore, \ch scales linearly with increasing number of cores, highlighting the efficiency of the parallelisation concept.

The implementation is not limited to inferring the parameters of $\Lambda$CDM using \camb and WMAP, but the application programming interface of \ch allows for the extension of its application to further cosmological probes and models using newly developed and self-contained Python modules.

In the appendix, we describe the installation of \ch and give a guide for the correct setup when using \camb and the WMAP likelihood.

%% file: appendixA.tex
\section{Download and Installation}\label{sec:downloadandinstallation}

The tarballs containing the most recent and stable version of \ch and the wrapper modules can be found at
\url{http://www.astro.ethz.ch/refregier/research/Software/cosmohammer}.

\ch relies on the following Python packages:
\begin{itemize}
	\item \emcee - affine invariant MCMC sampler 
	\item numpy - Numerical Python 
	\item mpi4py - Python wrapper for mpi (optional, only used if \ch is supposed to be distributed
	on multiple nodes) 
\end{itemize}

When additionally using the wrapped WMAP likelihood module, the WMAP data\footnote{\url{http://lambda.gsfc.nasa.gov/product/map/current/likelihood\_get.cfm}} needs to be accessible on the filesystem and CFITSIO\footnote{\url{http://heasarc.gsfc.nasa.gov/docs/software/fitsio/fitsio.html}} has to be installed. We tested \ch with Python 2.6, Python 2.7, numpy 1.6.2, \emcee 1.1.2, and mpi4py 1.3, but it is likely to work with earlier versions of these Python packages. For the compilation of the Fortran wrappers, a Fortran compiler and mkl libraries are required. The current distribution has only been tested with Intel's ifort compiler and mkl libraries, though.

To install the components the tarballs have to be extracted and the following commands have to be executed in the root directory of each module:

\begin{verbatim}
	% python setup.py build 
	% sudo python setup.py install 
\end{verbatim}
Every module comes with a Readme containing detailed information.

%% file: appendixD.tex
\section{Examples}\label{sec:examples}

We show how to use \ch to estimate cosmological parameters with \camb and WMAP likelihood. The listed source code is also available in the distributed tarball.

The import statements for both \ch and numpy have been omitted. We first define the initialisation by specifying the estimated start center, minimal and maximal value, and the start width using, for example, the values defined in Table \ref{tab:startDistribution}.

\lstset{language=Python, breaklines=true}

\begin{lstlisting}
#parameter start center, min, max, start width
params=np.array([[70, 40, 100, 3],
    [0.0226, 0.005, 0.1, 0.001],
    [0.122, 0.01, 0.99, 0.01],
    [2.1e-9, 1.48e-9, 5.45e-9, 1e-10],
    [0.96, 0.5, 1.5, 0.02],
    [0.09, 0.01, 0.8, 0.03]])
\end{lstlisting}

After instantiating the chain and passing the min and max parameter boundaries as follows, the chain will check if the proposed walker positions are within the boundaries before calling the modules.

\begin{lstlisting}
chain=LikelihoodComputationChain(
    min=params[:,1], 
    max=params[:,2])
\end{lstlisting}
We create an instance of the \emph{CambCoreModule} and add it to the previously instantiated chain. At this point it is possible to add other modules, e.g. further theory prediction modules.
\begin{lstlisting}
camb=CambCoreModule()
chain.addCoreModule(camb)
\end{lstlisting}

Next, we have to instantiate the WMAP likelihood computation module and then add the module to the chain. Here, further implementations of likelihood modules can be added. 

\begin{lstlisting}
wmapLikelihood=CmbWmapLikelihoodComputationModule() 
chain.addLikelihoodModule(wmapLikelihood)
\end{lstlisting}

Alternatively, we could create an instance of a \emph{CmbLikelihoodComputationChain} which automatically adds the previous modules.
\begin{lstlisting}
chain=CmbLikelihoodComputationChain(
	min=params[:,1], 
	max=params[:,2])
\end{lstlisting}

By calling the \emph{setup()} function the \camb and WMAP modules are initialised.

\begin{lstlisting}
chain.setup()
\end{lstlisting}

Finally, we create a \emph{CosmoHammerSampler} and pass the arguments. A \emph{walkersRatio} of 50 will launch $50 \times 6 = 300$ walkers, where six is the number of parameters sampled. By calling the \emph{startSampling()} function, \ch will first sample 250 iterations for burn-in and then run for another 250 iterations while writing the results to a file.

\begin{lstlisting}
sampler=CosmoHammerSampler(
	params=params, 
	likelihoodComputationChain=chain, 
	filePrefix="example", 
	walkersRatio=50, 
	burninIterations=250, 
	sampleIterations=250)

sampler.startSampling()
\end{lstlisting}

Further examples can be found in the distributed tarball.

%% file: appendixB.tex
\section{Estimation of autocorrelation}\label{sec:estimation_of_autocorrelation}

As mentioned in section \ref{sec:analysis}, we want to estimate the correlation function $C_{ff}(T)$ from a finite sample $\{\theta_t\}$. For parameter estimation, we are particularly interested in the case when $f(\theta_t) = \theta_t$ is the identity, as we want to estimate the mean of the marginals of our multidimensional posterior distribution (and maybe higher moments). For a single MCMC chain, the estimator for $C_{ff}(T)$ was introduced in equation \eqref{eq:cest} and for $f$ being the identity, it reads
\begin{equation*}
	\hat C(T) = \frac 1 {n - T} \sum_{t = 1}^{n - T}(X_t - \bar X)(X_{t + T} - \bar X),
\end{equation*}
where $X$ is one of the seven parameters of the likelihood and the estimator of the identity's autocorrelation function is denoted by $\hat C(T)$.

We now consider the case of an ensemble sampler with sample $\{X^i_t\}_{t \in 1, \cdots, n}^{i \in 1, \cdots, L}$ where $i$ is the walker and $t$ the iteration. \citet{Goodman2010a} propose the following procedure for estimating $C(T)$ of their ensemble sampler: Let $F_t = \frac 1 L \sum_{i = 1}^{L} X_t^i$ denote the ensemble average per iteration. The estimate $\hat C_{F}(T)$ of the correlation function of the process is then given by:
\begin{equation}
	\hat C_{F}(T) = \frac 1 {n - T} \sum_{t = 1}^{n - T}(F_t - \bar F)(F_{t + T} - \bar F).
	\label{eq:cFF}
\end{equation}

\begin{figure}[t]
	\centering
	\includegraphics[width=1.\linewidth]{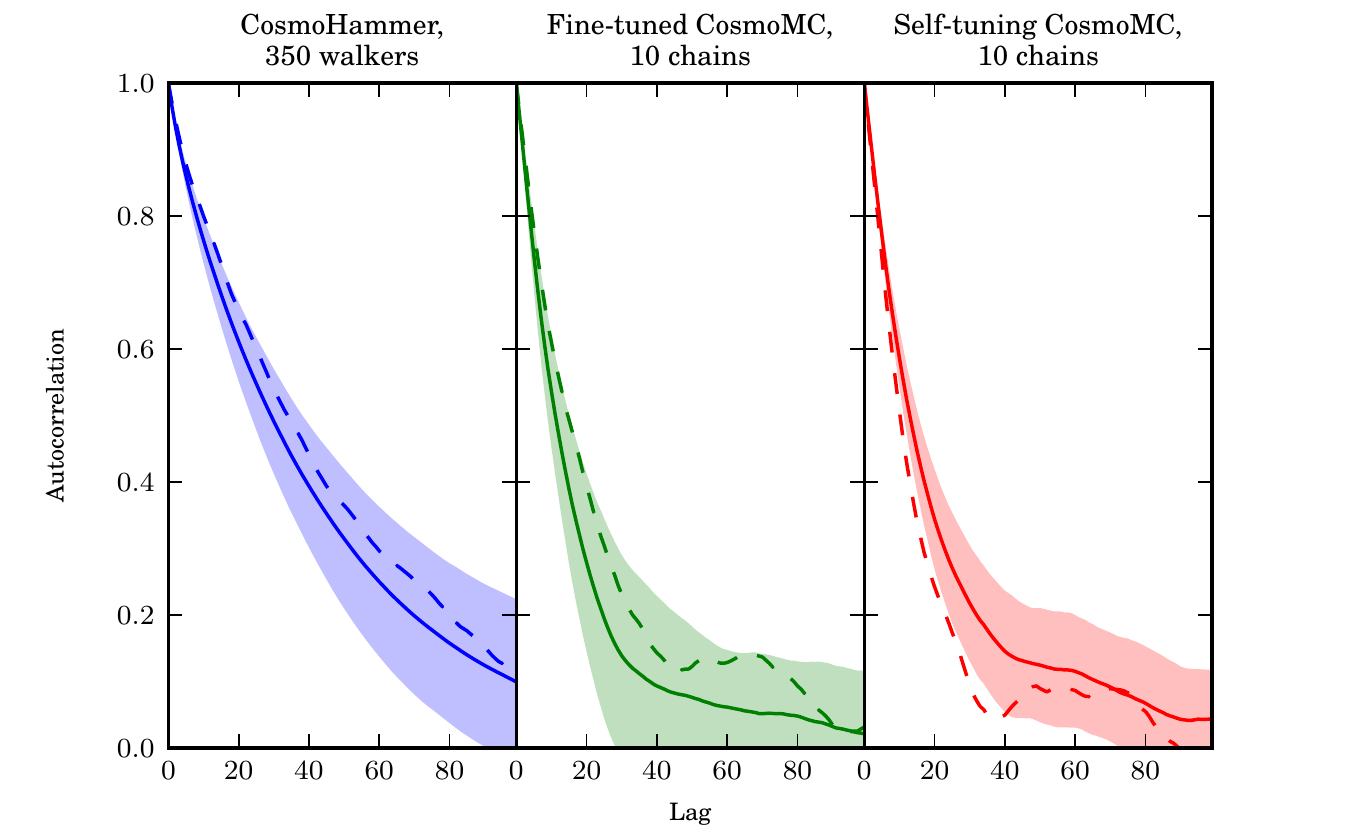}
	\caption{Autocorrelations of parameter $\Omega_{DM}h^2$ sampled by \ch, fine-tuned, and self-tuning \cosmomc with 3000 sequential steps, estimated by $\tilde C(T)$ (solid line) and $\hat C_{F}(T)$ (dashed line). The shaded regions depict the standard deviation of $\hat C(T)$ between the individual walkers or chains.}
	\label{fig:ensembleacor}
\end{figure}

In our analyses we found that large $n$ are needed for a stable estimator $\hat C_{F}(T)$ using this procedure. However, altering the calculations slightly did improve the stability while yielding results consistent with \eqref{eq:cFF}. Instead of calculating the autocorrelation of the average steps, we calculated $\hat C(T)$ per walker and averaged the results afterwards:
\begin{equation*}
	\tilde C(T) = \frac 1 L \sum_{i = 1}^L \frac 1 {n - T} \sum_{t = 1}^{n - T}(X^i_t - \bar X^i)(X^i_{t + T} - \bar X^i).
	\label{eq:ct}
\end{equation*}
The difference between the two approaches is visualised in Figure \ref{fig:ensembleacor} for parameter $\Omega_{DM}h^2$. It shows that $3,000$ steps are not sufficient for estimating the autocorrelation time reliably from a single chain, as the deviations between the different chains are still too big. Evaluating the autocorrelation time using the dashed line given by equation \eqref{eq:cFF} is hence not expected to yield reliable results. Averaging over a large number of walkers suppresses the fluctuations and improves the stability of the estimated autocorrelation time.

Even if the estimate for $C(T)$ is stable, it is non-trivial to deduce $\tau_{exp}$ and $\tau_{int}$ defined in equations \eqref{eq:tauexp} and \eqref{eq:tauint}. We used the algorithm proposed by Jonathan Goodman\footnote{\url{http://www.math.nyu.edu/faculty/goodman/software/acor/}} and implemented in Python by Dan Foreman-Mackey\footnote{\url{https://github.com/dfm/acor}} to estimate the integrated autocorrelation time. On the other hand, we evaluated the exponential autocorrelation time from a simple least-squares fit to $\tilde C(T)$ up to the maximal $T$ for which $\tilde C(T) < e^{-1}$ holds. 

As can be seen from Figure \ref{fig:ensembleacor}, the autocorrelation function resembles an exponential decay. We hence expect the estimated values of $\tau_{exp}$ and $\tau_{int}$ to coincide. Indeed, both $\tau_{exp}$ and $\tau_{int}$ converge to the same value, but the exponential autocorrelation time estimate is quite stable already at about $3,000$ iterations per walker (using 350 walkers), while the integrated autocorrelation has not yet converged. We hence decided to use $\tau_{exp}$ for estimating the relevant time-scales for our sample analysis.

Yet, it turns out that after analysing the autocorrelation time from $3,000$ iterations per walker and 350 walkers, we found in section \ref{sec:paramest} that only $250$ sequential steps are required for an error in the mean of 3.4\% of the standard deviation. Hence, more samples are needed to properly evaluate the error bars than for parameter estimation. Nevertheless, the estimate for $\tau_{exp}$ is typically of the right order even for small numbers of sequential steps. Rounding the result generously should hence suffice to get a rough upper bound on the error bars.

For our comparison of the samplers, we decided to consider runs with a large number of sequential steps in order to get stable estimates for $\tau_{exp}$. We used 10 independent chains with $3,000$ iterations each for the \cosmomc runs and $3,000$ iterations of $350$ walkers for the \ch. The results are stated in Table \ref{tab:tauexp}.